\documentstyle{article}%
\begin{document}

\title{CLePAPS: Fast Pair Alignment of Protein Structures
Based on Conformational Letters}
\author{Sheng Wang$^{1}$, Wei-Mou Zheng$^{1,2}$\\
${^1}${\it Institute of Theoretical Physics, Academia Sinica,
Beijing 100080, China}\\
$^{2}$To whom correspondence should be addressed}
\date{}
\maketitle

%\noindent{\bf Keywords:} pairwise structural alignments,
%conformational letters
\noindent E-mail addresses: zheng@itp.ac.cn
%\newpage

\begin{abstract}
Fast, efficient and reliable algorithms for pairwise alignment
of protein structures are in ever increasing demand for analyzing
the rapidly growing data of protein structures. CLePAPS is a tool
developed for this purpose. It distinguishes
itself from other existing algorithms by the use of conformational
letters, which are discretized states of 3D segmental structural
states. A letter corresponds to a cluster of combinations of the
three angles formed by $C_\alpha$ pseudobonds of four contiguous
residues. A substitution matrix called CLESUM is available to measure
similarity between any two such letters. CLePAPS regards an aligned
fragment pair (AFP) as an ungapped string pair with
a high sum of pairwise CLESUM scores. Using CLESUM scores as the
similarity measure, CLePAPS searches for AFPs by simple string comparison.
The transformation which best superimposes a highly similar AFP can be
used to superimpose the structure pairs under comparison. A highly scored
AFP which is consistent with several other AFPs determines an initial
alignment. CLePAPS then joins consistent AFPs guided by their
similarity scores to extend the alignment by several `zoom-in'
iteration steps.
A follow-up refinement produces the final alignment. CLePAPS does not
implement dynamic programming. The utility of CLePAPS is tested on
various protein structure pairs.
\end{abstract}
\medskip


\bigskip

\section{Introduction}

The comparison of protein structures has been an extremely important
problem in structural and evolutional biology. The detection of local
or global structural similarity between a new protein and a protein
with known function allows the prediction of the new protein's function.
Since protein structures are better conserved than amino acid
sequences, remote homology is detectable more reliably by comparing
structures. Structural comparison methods are useful for organizing
and classifying known structures, and for discovering structure
patterns and their correlation with sequences.

The common goal of all structure alignment methods is to identify a
set of residue pairs from each protein that are structurally similar,
or to find the optimal correspondence between the atoms in two
molecular structures. An exhaustive search for such atomic correspondence
between two structures is intractable, and various heuristics have been
developed. For example, to lower the dimensionality of the problem, DALI
identifies interaction patterns of fragment pairs,\cite{Holm1994,Holm1997}
VAST describes secondary structure elements (SSEs) as a set of
vectors,\cite{vast,vast1} while CE designates short aligned fragment
pairs (AFPs) of local structural similarities.\cite{ce}
There are several excellent reviews, e.g. \cite{rev00} and \cite{rev06}.
%Eidhammer et al. (2000) and Koehl (2006).

For a given correspondence of two point sets, finding the best rigid
transposition to superpose the correspondences can be easily done by
using a closed-form solution based on singular value decomposition.
\cite{Kabsch,svd} %(Umeyama, 1991).
When the transformation between the two sets is
given, the problem to find the correspondences (of $\epsilon$-congruence
at the maximal or average error tolerance $\epsilon$) is rather
straightforward. However, when aligning two protein structures, at
the beginning we know neither the transformation nor the correspondence.
A few methods, like DALI and CE, directly search for a good alignment.
Many methods start with an initial correspondence (seed matches), from
which the optimal transformation for the correspondence is determined.
The transformation is then used to update the correspondence. The
procedure of progressively building up larger correspondence is
iterated until the best correspondence is finally found. The methods
vary in the way of seed finding and correspondence updating. There
may be an optional follow-up refinement of alignments. A typical example
is ProSup.\cite{prosup} %(Lackner et al., 2000).

Protein structural alignment involves the geometric representation
of structures. In most cases, only the backbone of pseudobonds formed
by $C_\alpha$ atoms is considered. Coordinates of $C_\alpha$ atoms,
which change under translation and rotation in 3D space, are not
geometric invariants. Distances used by DALI are the intrinsic property
of a geometric object. The bending and torsion angles of pseudobonds, as
the chain counterparts of curvature and torsion of a smooth curve,
are also geometric invariants. In VAST, SSEs are replaced
by the vectors of their axes. This vector representation
speeds up the computation, but has low precision for structural
elements.

Many tools find
AFPs of local similarity as preliminary correspondences. Local
similarity is a necessary, but insufficient, condition for global
structural alignment. Structurally similar fragments
found in a pair of proteins form the basis objects for further
examination of their consistency in the spacial arrangement.
%required by the global alignment.
Consistent pieces then may be
joined to obtain a global alignment. Different methods use various
criteria and strategies for seed matching, consistency checking and
pieces merging. Generally, a stringent criterion for
local similarity would create less objects of seed matches, and hence
speed up the merging process. However, it would miss some substructures
constituting the final global alignment. On the other hand, due to
the insufficiencies of local similarity in the global alignment, too
loose conditions of local similarity would overload the later filtering
task. One has to balance sensitivity with specificity, and make a
compromise between efficiency and accuracy.

A way to represent structures is to use conformational alphabets,
which are discretized conformational states of certain fragment units
of protein backbones.\cite{rooman}--\cite{zheng-liu}
%\cite{rooman, park, edgoose, camproux, zheng-liu}
Our conformational alphabet of 17 letters is obtained by clustering
based on the distribution of the two
bending angles and one torsion angle formed by three pseudobonds
of the quadrupeptide unit. The description by conformational letters
provides a good balance between accuracy and simplicity, and converts a
3D structure to a 1D sequence of letters. Substitution matrices such as the
popular PAM and BLOSUM are essential to amino acid sequence
alignment algorithms. Without a conformational substitution
matrix the use of a conformational alphabet is very limited. In order
to implement fast structural comparison in terms of conformational
alphabets, we have derived a substitution matrix of conformational letters
called CLESUM from a representative pairwise aligned structure set
of the FSSP (families of structurally similar proteins) database of
Holm and Sander.\cite{holm1998a} It has been verified that
CLESUM aptly measures the similarity between conformational letter
states.\cite{zheng-liu}

Despite the existence of various pairwise structural alignment
algorithms, fast, efficient and reliable algorithms for pairwise
alignment are in ever increasing demand for analyzing the rapidly
growing data of protein structures. Here we report a tool called
CLePAPS developed for fast pairwise alignment of protein structures
by fully using our conformational alphabet and its substitution
matrix CLESUM. CLePAPS regards an AFP as an ungapped string pair with
a high sum of pairwise CLESUM scores. Using CLESUM scores as the
similarity measure, CLePAPS searches for AFPs by simple string comparison.
Taking highly similar AFPs as a pivot to determine the transformation
for superposition, CLePAPS collects consistent AFPs under the guide of
their similarity scores to extend the alignment.

\section{Methods}

CLePAPS uses a 3D structure coding of protein backbones consisting of
$C_\alpha$ pseudobonds. The flow chart of CLePAPS is shown in Fig.~1.

\subsection{Conformational alphabet}
Three contiguous $C_\alpha$ atoms determines two pseudobonds and a banding
angle between them. Four contiguous $C_\alpha$ atoms, say $a$, $b$, $c$ and
$d$, determine two such bending angles and a torsion angle which is the
dihedral angle between the two planes of triangles $abc$ and $bcd$.
By using a mixture model for the density
distribution of the three angles, the local structural states
have been clustered as 17 discrete conformational letters of a protein
structural alphabet. The centers ($\mu$), inverse covariance matrices
($\Sigma^{-1}$) and weights ($\pi$) of the clusters for these conformational
letters in the phase space spanned by the three angles $(\theta ,
\tau ,\theta ')$, are listed in Table 1.

%\newpage
\hspace{2cm}\parbox{7cm}{
\vspace{9cm}%\hspace{2cm}
\includegraphics{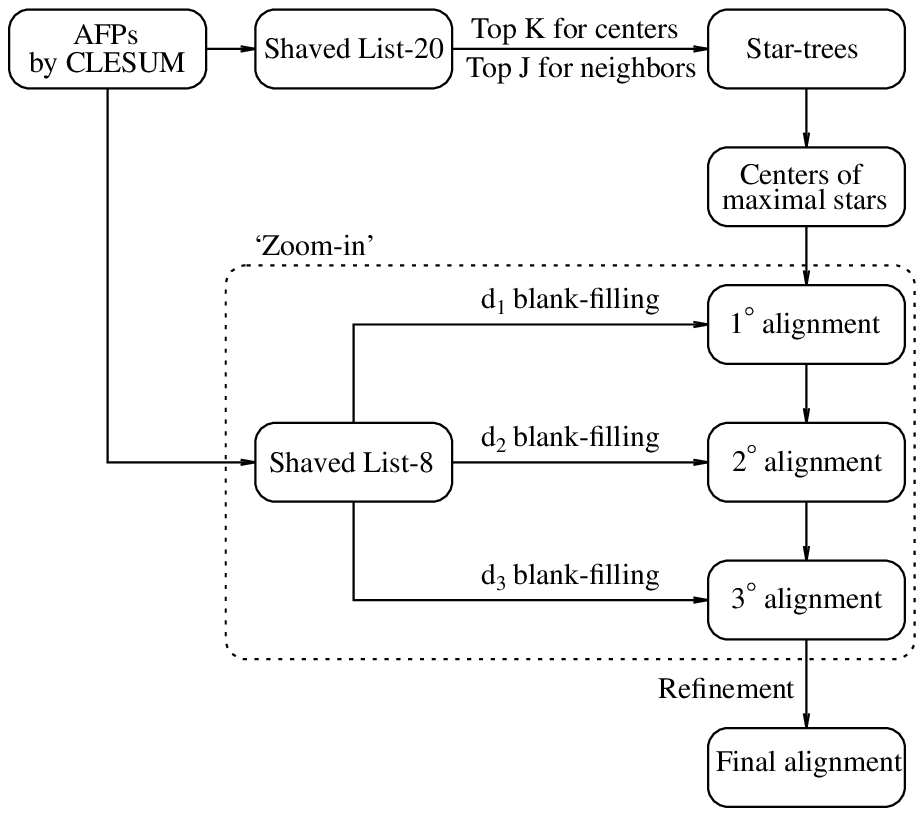}}%hscale=150 vscale=150}%%%
%\vspace{5cm}
%\begin{center}\parbox{10cm}
\begin{center}
{Fig.~1 %Illustration of CLePAPS.
Flow chart of the CLePAPS algorithm.}\end{center}
%\newpage

\begin{center}
\parbox{15cm}{Table 1. The 17 conformational states from the mixture
model. Angles are in radians.}\smallskip

\begin{tabular}{c|r|r|rrr|rrrrrr}\hline
&\multicolumn{1}{c|}{$\pi$}&\multicolumn{1}{c|}{$|\Sigma |^{-1/2}$}
&\multicolumn{3}{c|}{ ${\bf \mu}$ } &\multicolumn{6}{c}{${\bf\Sigma}^{-1}$}\\
State&& &\multicolumn{1}{c}{$\theta$}& \multicolumn{1}{c}{$\tau$}&
\multicolumn{1}{c|} {$\theta '$}&
\multicolumn{1}{c}{$\theta\theta$}& \multicolumn{1}{c}{$\tau\theta$}
& \multicolumn{1}{c}{$\tau\tau$}& \multicolumn{1}{c}{$\theta
'\theta$}
& \multicolumn{1}{c}{$\theta '\tau$}& \multicolumn{1}{c}{$\theta '\theta '$}\\
\hline
I&  8.2& 1881& 1.52&  0.83&  1.52& 275.4& $-$28.3&  84.3& 106.9& $-$46.1& 214.4\\ %(a1
J&  7.3& 1797& 1.58&  1.05&  1.55& 314.3& $-$10.3&  46.0&  37.8& $-$70.0& 332.8\\ %a2
H& 16.2&10425&1.55&  0.88&  1.55& 706.6& $-$93.9& 245.5& 128.9& $-$171.8&786.1\\ %a1
K&  5.9&  254&  1.48&  0.70&  1.43&  73.8& $-$13.7&  21.5&  15.5& $-$25.3&  75.7\\ %a'
F&  4.9&  105&  1.09& $-$2.72&  0.91&  24.1&   1.9&  10.9& $-$11.2&  $-$8.8&  53.0\\ %(b1
E& 11.6&  109&  1.02& $-$2.98&  0.95&  34.3&   4.2&  15.2&  $-$9.3& $-$22.5&  56.8\\ %b1
C&  7.5&  100&  1.01& $-$1.88&  1.14&  28.0&   4.1&   6.2&   2.3&  $-$5.1&  69.4\\ %b2
D&  5.4&   78&  0.79& $-$2.30&  1.03&  56.2&   3.8&   4.2& $-$10.8&  $-$2.1&  30.1\\ %(b2
A&  4.3&  203&  1.02& $-$2.00&  1.55&  30.5&   9.1&   8.7&   6.0&   5.7& 228.6\\ %rb
B&  3.9&   66&  1.06& $-$2.94&  1.34&  26.9&   4.6&   4.9&   9.5&  $-$5.0&  54.3\\ %r2
G&  5.6&  133&  1.49&  2.09&  1.05& 163.9&   0.6&   3.8&   2.0&  $-$3.7&  32.3\\ %ra
L&  5.3&   40&  1.40&  0.75&  0.84&  43.7&   2.5&   1.4&  $-$7.0&  $-$2.9&  34.5\\ %(ra
M&  3.7&  144&  1.47&  1.64&  1.44&  72.9&   2.1&   4.8&   1.9&  $-$7.9&  72.9\\ %a+
N&  3.1&   74&  1.12&  0.14&  1.49&  25.3&   3.2&   3.1&   9.9&   0.9&  83.0\\ %r1
O&  2.1&  247&  1.54& $-$1.89&  1.48& 170.8&  $-$0.7&   3.7&  $-$4.1&   3.1&  98.7\\ %a-
P&  3.2&  206&  1.24& $-$2.98&  1.49&  48.0&   8.2&   7.3&  $-$4.9&  $-$6.6& 155.6\\ %(r2
Q&  1.7&   25&  0.86& $-$0.37&  1.01&  28.4&   1.5&   1.2&   3.4&   0.1&  19.5\\ %r'
\hline\end{tabular}\end{center}
\smallskip

To use our structural codes directly for the structural comparison,
a score matrix similar to the BLOSUM for amino acids is desired.
Using the alignments for representative structures in the database
FSSP, which contains 2,860 sequence families representing 27,181
protein structures, we have constructed a substitution
matrix called CLESUM for the conformational letters.
The structures of the representative set are converted to their
structural code sequences. All
the pair alignments of the FSSP for the proteins with a
sufficient similarity in the representative set are collected
for counting aligned pairs of conformational letters. The total
number of letter pairs is 1,284,750. An entry of the matrix is
the log-ratio of the observed frequency of the aligned
corresponding pair to the expected frequency from a random
alignment simply by chance. The substitution matrix is derived
in the same way as BLOSUM was obtained (without clustering).
The matrix is shown in Table 2, where a scaling factor of 20
instead of 2 is used to show more details (corresponding to
$0.05$ bit units).

To the best of our knowledge, CLESUM is the first substitution matrix
directly derived from structure alignments for a conformational
alphabet. (Matsuda et al. \cite{3dcode} %(1997)
introduced a backbone encoding which is a uniform partition of the 2D
phase space of one bending and one torsion angles by mapping the phase
space into the surfaces of a icosahedron; their substitution matrix is
given by the cosine of the angles between the corresponding normals of
the icosahedron. This matrix measures a geometrical similarity while
CLESUM reflects an `evolutionary' similarity.)

\begin{center}%}\end{center}\smallskip
\parbox{15cm}{Table 2. CLESUM: The conformation letter substitution
matrix (in 0.05 bit units).\\
}\end{center}\vspace{-.5cm}

{\small\tt \begin{tabular}{c rrrrrrrrrrrrrrrrr}\hline
J& 37&    &    &    &    &    &    &   &   &   &   &   &   &   &   &   &   \\
H& 13&  23&    &    &    &    &    &   &   &   &   &   &   &   &   &   &   \\
I& 16&  18&  23&    &    &    &    &   &   &   &   &   &   &   &   &   &   \\
K& 13&   5&  21&  49&    &    &    &   &   &   &   &   &   &   &   &   &   \\
N& -2& -34& -11&  28&  90&    &    &   &   &   &   &   &   &   &   &   &   \\
Q&-44& -87& -62& -24&  32&  90&    &   &   &   &   &   &   &   &   &   &   \\
L&-32& -62& -41&  -1&   8&  26&  74&   &   &   &   &   &   &   &   &   &   \\
G&-21& -51& -34& -13&  -8&   8&  29& 69&   &   &   &   &   &   &   &   &   \\
M& 16&  -4&   1&  12&   7&  -7&   5& 21& 61&   &   &   &   &   &   &   &   \\
B&-57& -96& -74& -50& -11&  12& -12& 13&-13& 51&   &   &   &   &   &   &   \\
P&-34& -60& -49& -36&  -3&   7& -12&  5&  8& 42& 66&   &   &   &   &   &   \\
A&-23& -45& -31& -19&  10&  16& -11& -6& -2& 20& 35& 73&   &   &   &   &   \\
O&-24& -55& -34&   5&  15& -13&  -4& -1&  5&-12&  4& 25&104&   &   &   &   \\
C&-43& -77& -56& -33&  -5&  29&   0& -4&-12&  7&  4& 13&  3& 53&   &   &   \\
E&-93&-127&-108& -84& -43&  -6& -21&-22&-47& 15& -5&-25&-48&  3& 36&   &   \\
F&-73&-107& -88& -69& -32&   3& -16& -5&-33&  7&  0&-20&-30& 20& 26& 50&   \\
D&-88&-124&-105& -81& -44&  14& -22&-31&-49& 13&-10&-17&-42& 21& 22& 21& 52\\
  & J &  H &  I&   K & N &   Q &  L & G & M & B & P & A & O & C  &E & F & D\\
\hline\end{tabular}}
\medskip

\subsection{Finding AFPs of high CLESUM similarity scores}

Suppose that the pair of structures to be aligned is
$P$ and $P'$ with $P$ being the
shorter. The coordinates $\{{\bf r}_{i}\}$ and $\{{\bf r}'_{j}\}$
of $C_\alpha$ atoms of the two proteins are converted to the sequences
$S$ and $S'$ of conformational letters, respectively. Since each letter
corresponds to a quadrupeptide unit, the length of $S$ ($S'$) is
shorter than that of $P$ ($P'$) by three. By convention, we assign
the first letter to the third residue, the second to the fourth, and
so on, until finally the last letter is assigned to the last residue but one.

Consider two fragments of the same length $l$, one of which starts at residue $i$
of $P$ and the other at $j$ of $P'$. The local  structural similarity
of the fragment pair may be measured by
\begin{equation}
\sigma = \sum_{k=0}^{l-1} M(s_{i+k},s'_{j+k}),
\end{equation}
where $M(a,b)$ is the $(a,b)$-entry of the CLESUM, and $s_i$ and $s'_j$
are the conformational letters of corresponding residues. Here we have kept
the same index for a residue and its conformational letter. Setting a threshold
$T$, if the pair score $\sigma \geq T$, we call the pair an AFP, which
defines a correspondence ($l$ residue duads).
Comparing each string of $S$ of length $l$ with those of $S'$, we find all
the AFPs. When an AFP contains residues of SSEs at its ends, shifts
of the AFP often also have a high similarity score, hence they also form AFPs. To
remove such redundancy, we keep only the one with the highest score
among the nearby AFPs which are shifts of each other and which share common duads.
A width $w$ is set to restrict the maximum overlap for this shaving. After
shaving we have a reduced list of the representative AFPs. We sort
the list in descending order of scores.
Usually, a small $l$ and a low $T$ will result in a long list of AFPs.

For a given long enough AFP, we can find a rigid transformation to superpose
its two members and make the spatial deviation of its duad $C_\alpha$
atoms very small. Since an AFP is determined only by local similarity, a
superposition valid for one AFP need not be valid for another. We define
the spatial distance or separation between two members of a certain AFP
under a given transformation by
\begin{equation}
\delta = \max_{({\bf r}_i,{\bf r}'_{i'})\in {\rm AFP}}
\{|x_i-x'_{i'}|,|y_i-y'_{i'}|,|z_i-z'_{i'}|\},
\end{equation}
where $({\bf r}_i,{\bf r}'_{i'})$ is a duad of the AFP after transformation,
and $(x,y,z)$ denotes the 3D coordinates of ${\bf r}$. A small separation
$\delta$ implies a good superposition of the two AFP members.

\subsection{The greedy `zoom-in' strategy}

There is no clearly defined unique way to evaluate the quality of protein
structure alignments. We adopt the standard of ProSup: the goal is to
maximize the number $N_e$ of structurally equivalent residues subject to
a fixed Euclidean distance cutoff $d_0$ for judging correspondence between a
residue pair and a minimal aligned segment size $\rho$.

To balance speed with accuracy, we generate two lists of AFPs, one for
$l=20$ with threshold $T_{20}=350$, and the other for $l=8$ with
$T_8=0$. Any two helices are locally similar. Length 20 will exclude many
such purely local coincidence. Length 8 is necessary for including most
significant aligned pieces. We denote them as List-20 and List-8, respectively.
The two lists can be generated in a single run. We expect that a significant
alignment should contain at least one AFP of length around 20.
Initial primary correspondences will be taken from the top $K$, say top
ten, AFPs of the sorted List-20. If the list size is less than 10,
`top ten' means all.

Once an AFP is chosen, the transformation optimal to the AFP may be used
to superimpose the two proteins. The separation $\delta$ of any
AFP under the transformation can then be calculated. Some AFPs are
consistent with each other. That is, under the transformation optimal to
one of the AFPs, which is referred to as the center AFP or simply the
center, some other AFPs will have a small separation $\delta$. By thinking
in terms of graph theory, AFPs are nodes, and the center is the center node.
At a fixed threshold
$d$ for $\delta$, if the separation of an AFP is smaller than $d$, an
edge is linked between the node of that AFP and the center. The
center and its linked neighbors form a star tree or star. We define
the size of a star as the total number of its nodes. Taking
each of the top $K$ AFPs as a center, we find its neighbor nodes in
the top $J$ AFPs of the sorted List-20, where $J\geq K$. The stars
really used by CLePAPS are subject to a further restriction: for a given
center, we search the sorted List-20 successively from the top for
neighbors of the center, and add a new neighbor AFP only when it does
not overlap with any existing neighbor AFPs. In this way, we obtain $K$
restricted stars. We sort them first by their sizes, and then by the 
similarity score $\sigma$ of their centers in descending order.
We remove the stars whose centers are a neighbor of the first star.
Then, we examine the next star, and remove the stars associated with
its neighbors, and so on, until all stars are
examined. Only the centers of the retained stars will be taken as an
initial alignment seed. The effect of this star removal is twofold:
removing seed redundancy and selecting the seeds which better reflect
the global consistency.

The extension of an initial seed alignment is mainly done by 
blank-filling of the AFPs from List-8 which are consistent with the seed.
Blanks are residue positions not included in an existing
correspondence. The transformation optimal to the seed AFP need not be
globally optimal. We use a multi-step `zoom-in' strategy, starting with a
low precision to avoid local trapping. We first use a large cutoff $d_1$,
say 8\AA, as the consistency criterion. That is, we add only the
AFPs with $\delta <d_1$
to the existing correspondence. The procedure of blank-filling is greedy.
The AFPs with a higher $\sigma$ have a priority to be filled.
We examine the top half of the AFPs in the sorted List-8 one by one from the top.
If none of the residues of an AFP is contained in the existing
correspondence, we calculate its separation $\delta$. If $\delta <d_1$,
we add the AFP to the correspondence.
%At each position
%for blank-filling we examine only up to $k$ AFPs from List-8, and add
%the one with the smallest $\delta$ if $\delta <d_1$. If a later
%blank-filling conflicts with an early one, only the early one is kept.
When blank-filling is fulfilled, the transformation optimal
to the enlarged correspondence is determined to update the superposition
of the two proteins. In the next run of blank-filling, cutoff $d_1$ is reduced
to $d_2$, and five-sixths of the AFPs in List-8 are examined. In a third run,
$d_2$ is further reduced to $d_3$, and the whole list is examined. Usually,
three runs of iteration are enough for obtaining a full alignment.

There are mainly two ways to update the correspondence. One is to keep
the existing duads and add new ones. The other is to re-start with
an empty correspondence and then fill in blanks with AFPs from List-8.
The latter strategy is used in CLePAPS. In the
final polishing stage, the AFPs which have only a limited overlap with
the existing correspondence can also be used for blank-filling.

We speed up computation by means of marking. At the beginning, all AFPs
in List-8 are identified as `unmarked'. If an AFP has no two contiguous
residue pairs whose coordinate differences are both
smaller than $d_i$, it will be marked as
`invalid', and then will never be examined again.
%(* correspondence: 1D-array indexed by positions of protein 1. entry:
%a positive number if duad exits, -1 if not suitable for filling; $-2 - $
%checked times for position suitable for filing. allow an overlap up to 4
%when blank-filling (check the entry at beginning of AFP+4)? )

\subsection{Refinement by elongation and shrinking}

After blank-filling we obtain an alignment usually as disjunct
pieces. Due to the finite size of AFPs and the redundancy removal by
shaving, it is possible that an aligned piece can be
elongated near its ends. Assume that the residue duad $(i,i')$ is at one end
of an AFP. Thus, either $(i+1,i'+1)$ or $(i-1,i'-1)$ will be an outer residue
pair. If the Euclidean distance of the pair is smaller
than $d_0$ we elongate the aligned piece by joining the residue duad
to it. More nearby residue pairs can be further examined for
elongation. On the other hand, depending on the relative quantities
of $d_0$ and $d_3$, the Euclidean distance between some residue duads on the
aligned pieces may be greater than $d_0$. We remove such pairs
from the alignment (as a shrinkage of the AFPs). Although elongation and
shrinking may be conducted subsequently for every complete run of
blank-filling, CLePAPS conducts elongation and
shrinking only after the final run.

A finetuning version of refinement is as follows. In the final run of
blank-filling, position confliction in protein $P'$
is ignored so one residue of protein $P$ may correspond to several
residues of protein $P'$. We sort in ascending order of the root
mean square deviations (RMSDs) all the AFPs which are candidates
for filling, then add these AFPs into an empty correspondence
list one at a time and avoid any position
confliction. We perform possible elongation of each filled AFP to
enlarge the correspondence. For any unfilled gap in
protein $P$, we search List-8 from the top. If an AFP which covers
the gap is found, we calculate the Euclidian distance of residue
pairs whose positions coincide with the gap. When the distance of
a pair is smaller than $d_0$, we check also the distances of nearby
pairs by shifting protein $P'$, and add
the pair with the smallest distance to the correspondence.

A filter for a minimal aligned segment length $\rho$ may
be finally applied. A further iteration of transformation
would additionally improve the quality of the alignment. Once a global
alignment is accomplished, the total number $N_e$ of equivalent
residue pairs and the RMSD of the
alignment are calculated as quality indicators. Despite the star removal
two alignments generated from two star centers may still be very similar.
We compare entries of the rotation matrices. If the greatest difference
between two corresponding matrix elements (or relative difference) is
above $\epsilon$, say 0.1,
the two alignments are regarded as identical. A more careful criterion
is the number of coincident residue duads, which is used by ProSup.

Structure comparison often yields several distinct alignments as
multiple solutions. The existence of alternative alignments is
mainly due to structure repeats at different levels ranging from
secondary structure, supersecondary structure %, internal repeats
to domains.\cite{prosup} %ProSup
Another source is the domain move. CLePAPS often reports several
alignments and ranks them according to their $N_e$.

\section{Results}

\subsection{Finding AFPs in the test case: phycocyanin versus colicin A}
This test case was thoroughly studied by CE for tuning its operation
parameters. The two proteins have PDB codes 1colA and 1cpcL. Here letters
A and L are chain identifiers. The former is classified as a `membrane and cell
surface protein', and the latter as an `all alpha protein' by SCOP.\cite{scop}
Their lengths are 204 and 172, respectively. Only 197 residues of 1colA 
are given with their coordinates.

A great proportion of both the protein structures is helices. It is
expected that the number of AFPs should be large. Numbers of AFPs
found at two different lengths $l$ and different CLESUM score
thresholds $T$ before and after shaving are shown in Table 3.
The maximum overlap for shaving is set to be $i\times l/2$, and three
overlap ranges for shaving are examined. From the table it is clearly
seen that the number of AFPs drops significantly for long $l$
and high $T$, and shaving dramatically reduces the number of
AFPs. Finally, in CLePAPS the maximum overlap is set to be 20
for List-20 and 4 for List-8.

\begin{center}%}\end{center}\smallskip
\parbox{15cm}{Table 3. Numbers of AFPs found at different lengths
$l$ and CLESUM score thresholds $T$ before and after shaving. $N_i$
denotes the number of AFPs after shaving at overlap range $i\times l/2$.
}%\end{center}\vspace{-.5cm}

\begin{tabular}{cc r rrr}\hline
&&\multicolumn{1}{c}{before}&\multicolumn{3}{c}{after shaving}\\
$l$& $T$& $N_0$& $N_1$& $N_2$& $N_3$\\
\hline
  & 500 & 21  & 7  &6  &6\\
20& 400& 164 & 45 &42 &41\\
   &350& 381  &93 &87 &85\\
   &300 &717& 156&140&135\\
\hline
  &200  &513  &239& 200& 195\\
  &150 &4273 &1512& 1093& 953\\
8& 100 &7753 &2462 &1616 &1245\\
   &\phantom{1}50&10350& 3161 &1960 &1418 \\
   & \phantom{10}0&13207& 3942 &2316 &1547 \\
 \hline
\end{tabular}\\ \end{center}

We have examined the pair 1colA: 1cpcL and many other pairs to optimize
the operation parameters of CLePAPS. We use the default values of these
parameters as shown in Table 4.

\begin{center}%}\end{center}\smallskip
\parbox{15cm}{Table 4. Default parameters of CLePAPS.
}%\end{center}\vspace{-.5cm}

\begin{tabular}{cc l}\hline
Symbol& Value& \multicolumn{1}{c}{Meaning}\\
\hline
$l_l$& 20& length of long AFPs\\
$T_l$& 350& similarity threshold for long AFPs\\
$K$& 10& number of long AFPs used as seed candidates\\
$J$& 50& number of long AFPs for building a star-tree\\
$l_s$& 8& length of short AFPs for blank-filling\\
$T_s$& 0& similarity threshold for short AFPs\\
%$k$& 2000& number of short AFPs examined for blank-filling\\
$\rho$& 4& minimum length of aligned fragments\\
$d_0$& 5\AA& distance cutoff for evaluating overall alignment\\
$d$ & 10\AA &separation threshold for star construction\\
%$d^+$& 12\AA& separation bound for masking\\
$d_1$& 8\AA& separation cutoff for blank-filling in first run\\
$d_2$& 6\AA& separation cutoff for blank-filling in second run\\
$d_3$& 5\AA& separation cutoff for blank-filling in third run\\
$\epsilon$& 0.1& maximal difference for rotational matrix \\
&& entries of two `identical' alignments\\
\hline
\end{tabular}\\ \end{center}

\subsection{The Fischer benchmark test} %Comparison with popular algorithm DALI}

A well-known comprehensive test set for assessing the performance
of fold recognition methods is the benchmark of Fischer et al.,
which contains 68 pairs of proteins.\cite{fischer} All pairs of the
set are known to be structurally similar, but they have low sequence
identity, ranging from 8\% to 31\% with an average of 18.6\% and a
standard deviation of 4.4. This set covers a wide range of protein
families. We test the benchmark with CLePAPS. The
results of the alignment are summarized in Table 5. Although AFPs
reflect mainly the local similarity, the construction of star trees
helps us to select seed AFPs for pivoting superposition. We rank
seeds first according to the sizes of their center-stars, and then
according to the similarity scores $\sigma$ of the center seeds.
The last column of the table is the rank of the center seeds from
which the optimal alignments are derived.

\begin{center}%}\end{center}\smallskip
\parbox{15cm}{Table 5. Test of CLePAPS on the Fischer benchmark.
ID: protein PDB ID and an optional fifth letter for chain index;
$L$: protein length; Rank: the rank of the center star from whose center
the optimal alignment is generated; $N_e$: the number of aligned
residues; rmsd: RMSD of the alignment. Nine pairs whose optimal
alignments are not from the rank-1 seed AFPs are indicated
with a superscript in the last column for ranks. Superscripts $a$, $b$ and
$c$ correspond to three groups of the nine pairs, see text.
}%\end{center}%\vspace{-.5cm}
\smallskip

\begin{tabular}{lrlr rrrc}\hline
\multicolumn{4}{c}{Protein pair}&\multicolumn{1}{c}{DALI}
&\multicolumn{1}{c}{DALI-core}&\multicolumn{2}{c}{CLePAPS}\\
ID& $L$& ID& $L$& $N_e$/rmsd & $N_e$/rmsd& $N_e$/rmsd& Rank \\
\hline
%ID    L   ID    L  rank $N_e$ rmsd
1mdc & 133& 1ifc & 131&      -  & -       & 127/1.87& 1\\
1npx & 447& 3grs & 461& 395/3.45& 347/2.40& 335/2.38& 1\\
1onc & 104& 7rsa & 124&  97/1.86&  91/1.49&  91/1.56& 1\\
1osa & 148& 4cpv & 108&  67/1.43&  67/1.43&  67/1.43& 2$^a$\\ %1
1pfc & 111& 3hlaB&  99&  88/2.83&  74/2.14&  77/2.29& 1\\
2cmd & 312& 6ldh & 329& 286/2.52& 269/2.05& 271/2.09& 1\\
2pna & 104& 1shaA& 103&  92/2.62&  85/2.14&  85/2.20& 1\\
1bbhA& 131& 2ccyA& 127& 125/2.02& 121/1.86& 122/1.81& 1\\
1c2rA& 116& 1ycc & 108&  96/1.62&  95/1.53&  95/1.49& 1\\
1chrA& 370& 2mnr & 357& 347/1.88& 340/1.73& 340/1.73& 1\\
1dxtB& 147& 1hbg & 147& 135/2.04& 128/1.74& 137/1.95& 1\\
2fbjL& 213& 8fabB& 214& 194/2.30& 186/1.91& 186/1.90& 1\\
1gky & 186& 3adk & 194& 154/2.97& 129/2.43& 122/2.44& 1\\
1hip &  85& 2hipA&  71&  67/1.81&  66/1.71&  62/1.43& 1\\
2sas & 185& 2scpA& 174& 168/3.58& 131/2.53& 134/2.49& 1\\
1fc1A& 206& 2fb4H& 229& 175/8.28& 101/2.01& 117/2.23& 1\\
2hpdA& 457& 2cpp & 405& 374/3.47& 319/2.73& 307/2.64& 1\\
1aba &  87& 1ego &  85&  72/2.19&  68/1.73&  68/1.73& 1\\
1eaf & 243& 4cla & 213& 174/2.62& 161/2.25& 165/2.35& 1\\
2sga & 181& 5ptp & 222& 147/2.71& 128/1.87& 139/2.02& 1\\
2hhmA& 278& 1fbpA& 316& 224/2.85& 195/2.11& 203/2.27& 1\\
1aaj & 105& 1paz & 120&  80/1.67&  79/1.56&  79/1.61& 1\\
5fd1 & 106& 1iqz &  81&  57/2.62&  47/1.95&  51/1.53& 5$^a$\\%2 5fd1 cath a+b,2-layer sandwich,ab plaits; pf00037*2. scop Ferredoxin-like%1iqz  pf02990. 48/2.1; 45/1.9
1isuA&  62& 2hipA&  71&  58/2.28&  51/1.55&  54/1.98& 1\\
1gal & 581& 3cox & 500& 401/3.05& 338/2.19& 342/2.18& 1\\
1cauB& 184& 1cauA& 181& 162/2.18& 153/1.80& 154/1.82& 1\\
1hom &  68& 1lfb &  77&  56/1.95&  52/1.40&  55/1.65& 1\\
1tlk & 103& 2rhe & 114&  89/2.02&  81/1.41&  86/1.70& 1\\
2omf & 340& 2por & 301& 261/2.68& 231/2.06& 228/2.06& 1\\
1lgaA& 343& 2cyp & 293& 261/2.44& 235/1.84& 241/1.92& 1\\
1mioC& 525& 2minB& 522& 412/3.61& 353/2.59& 361/2.63& 1\\
 \hline
\end{tabular}\end{center}
%\noindent (Table 5, continued)
\smallskip

Amongst 68 protein pairs, the centers of 59 (87\%) rank-1 star
trees of List-20 lead to optimal alignments.
The nine protein pairs whose optimal alignments do not correspond
to a rank-1 star tree form three groups $a$, $b$ and $c$. Four pairs in
group $a$ have structural repeats, and the rank-1 stars do correspond
to one of the multiple best choices although their $N_e$ are relatively
lower. For the two pairs in group $b$, the alignments from the rank-1 stars
are similar to the optimal. For one pair of the two, 1dsbA: 2trxA, the
rank-1 center AFP is consistent with the optimal alignment.
However, it leads to only partial alignment due to local trapping. An
extra `zoom-in' step with cutoff 10\AA\ is able to obtain the optimal
alignment from that AFP. The other three of group $c$ (1aep: 256bA,
1rcb: 2gmfA and 1bgeB: 2gmfA) belong to a `four-helical up-and-down bundle'.
The alignments of DALI consist of almost just helical regions; DALI
and CE have nothing in common for the alignment between 1aep and 256bA.
It seems that the evidence of local similarity to support the global
alignments of these three pairs is not as strong as for other pairs in the
benchmark. We
shall come back to the pair 1bgeB: 2gmfA in the next subsection. It seems
that the loss of sensitivity by keeping only the rank-1 star and
obtaining alignment only from its center is rather limited.

\begin{center}
\begin{tabular}{lrlr rrrc} %\hline
\multicolumn{8}{l}{(Table 5, continued)}\\
\hline
\multicolumn{4}{c}{Protein pair}&\multicolumn{1}{c}{DALI}
&\multicolumn{1}{c}{DALI-core}&\multicolumn{2}{c}{CLePAPS}\\
ID& $L$& ID& $L$& $N_e$/rmsd & $N_e$/rmsd& $N_e$/rmsd& Rank \\
\hline
4sbvA& 199& 2tbvA& 287& 162/2.09& 154/1.47& 154/1.47& 1\\
8i1b & 146& 4fgf & 124& 118/2.47& 110/1.96& 110/2.08& 2$^a$\\ %3
1hrhA& 125& 1rnh & 148& 114/1.98& 105/1.40& 105/1.37& 1\\
1mup & 157& 1rbp & 174& 140/2.92& 123/2.12& 120/2.10& 1\\
1cpcL& 172& 1colA& 197& 114/3.61&  99/3.02& 104/2.98& 1\\
2ak3A& 226& 1gky & 186& 149/3.00& 120/2.35& 118/2.41& 1\\
1atnA& 373& 1atr & 383& 292/3.00& 257/2.34& 258/2.32& 1\\
1arb & 263& 5ptp & 222& 189/2.89& 164/2.10& 159/2.13& 1\\
2pia & 321& 1fnb & 296& 216/2.52& 196/1.98& 195/2.12& 4$^b$\\ %4
3rubL& 441& 6xia & 387& 206/4.14& 142/3.30& 154/2.94& 2$^a$\\ %5
2sarA&  96& 9rnt & 104&  71/3.18&  58/2.46&  61/2.66& 1\\
3cd4 & 178& 2rhe & 114&  94/2.60&  88/1.49&  89/1.63& 1\\
1aep & 153& 256bA& 106&  74/1.78&  73/1.63&  84/2.83& 4$^c$\\ %6 ce 94/4.1=5.3, da 74/1.8=5.7
2mnr & 357& 4enl & 436& 285/3.43& 237/2.72& 227/2.71& 1\\
1ltsD& 103& 1bovA&  69&  67/1.92&  66/1.82&  66/1.91& 1\\
2gbp & 309& 2liv & 344& 260/6.75& 109/2.37& 140/2.51& 1\\
1bbt & 186& 2plv & 288& 168/2.62& 159/2.31& 150/2.19& 1\\
2mtaC& 147& 1ycc & 108&  80/2.10&  79/1.99&  71/1.94& 1\\
1tahA& 318& 1tca & 317& 188/2.47& 178/2.15& 173/2.22& 1\\
1rcb & 129& 2gmfA& 121&  82/3.32&  67/2.32&  82/2.77& 2$^c$\\ %7 da=5.8, ce 104/4.3=4.4
1sacA& 204& 2ayh & 214& 133/3.03& 118/2.67& 132/2.79& 1\\
1dsbA& 188& 2trxA& 109&  82/2.77&  77/2.04&  77/2.01& 2$^b$\\ %8
1stfI&  98& 1molA&  94&  85/1.92&  85/1.92&  85/2.29& 1\\
2afnA& 331& 1aozA& 552& 248/2.56& 231/2.23& 223/2.24& 1\\
1fxiA&  96& 1ubq &  76&  60/2.58&  55/2.30&  50/2.30& 1\\
1bgeB& 159& 2gmfA& 121&  94/3.33&  79/2.22&  82/2.41& 2$^c$\\ %9 ce 107/3.9=4.1, da=6.6
3hlaB&  99& 2rhe & 114&  75/3.03&  63/2.32&  65/2.30& 1\\
3chy & 128& 2fox & 138& 103/3.04&  91/2.59&  86/2.76& 1\\
2azaA& 129& 1paz & 120&  81/2.26&  78/1.92&  78/2.30& 1\\
1cew & 108& 1molA&  94&  81/2.46&  76/2.14&  78/1.98& 1\\
1cid & 177& 2rhe & 114&  97/3.15&  82/2.03&  87/2.21& 1\\
1crl & 534& 1ede & 310& 211/3.47& 168/2.49& 169/2.65& 1\\
2sim & 381& 1nsbA& 390& 292/3.26& 240/2.50& 248/2.61& 1\\ %1sim->2sim, new version
1ten &  89& 3hhrB& 195&  86/1.91&  84/1.73&  84/1.71& 1\\
1tie & 166& 4fgf & 124& 114/3.06&  97/2.18& 100/2.26& 1\\
2snv & 151& 5ptp & 222& 131/3.09& 113/2.38& 118/2.49& 1\\
1gp1A& 183& 2trxA& 109&  97/3.70&  74/3.40&  85/2.22& 1\\
 \hline
\end{tabular}
\end{center}

%\begin{center}
%\parbox{15cm}{%From Ye JP's Table 3 (``bone\_ali.pdf").
%Comparison of CLePAPS with VAST, DALI and CE on some protein pairs
%from the benchmark of Fischer et al. \\}
%%LOCK and DALI on the Fischer benchmark. \\ }
%
%\begin{tabular}{lc lc rc rc rc}\hline
%\multicolumn{4}{c}{Protein pair}&\multicolumn{2}{c}{YE}&
%\multicolumn{2}{c}{LOCK}& \multicolumn{2}{c}{DALI}\\
%PDB&$L$& PDB&$L$& $N_e$&{\small RMSD}&$N_e$&{\small RMSD}&
%$N_e$&{\small RMSD}\\
%\hline
%1mdc& 133& 1ifc   &131 &128 &1.84 &33 &1.91 & -& -\\
%1npx& 447& 3grs   &461 &384 &2.65 &32 &1.62 &395&3.50\\
%5fd1& 106& 1iqz   &\phantom{1}81  & 44 &2.87 &46 &1.48 &57 &2.60\\
%1aep& 153& 256bA  &106 & 55 &2.28 &49 &1.74 &74 &1.80\\
%\hline
%\end{tabular}\\ \end{center}

Indeed, with
the aid of the star tree construction, the AFPs in terms of CLESUM
scores are efficient for generating protein structure alignment.
The selection of a seed for initial correspondence plays an essential
role in the quality of the final alignment.
In most cases (41 of 68) the AFP of List-20 with the highest $\sigma$
is the one from which the optimal alignment is derived.
It appears that $K=10$ is
enough to include the seed leading to the optimal alignment. A large
value of $J$ for construction of a star tree is favorable for picking
up a right seed. The size of a star tree is small for a small
$J$. When the structures to be compared are large or contain repeats,
usually a large $J$ is required for star tree construction. For
example, the pair 2sim: 1nsbA is a highly repeated fold `6-bladed
beta-propeller' with lengths 381 and 390, respectively. At
$J=30$, every star tree is of size 1, consisting of only the center
itself. In CLePAPS we use $J=50$; increasing $J$ from 30 to 50 does not
add much extra computational cost.

In the table we have also listed $N_e$ and RMSD of DALI alignments. It
is difficult to compare them directly with those from CLePAPS. To make
a close comparison, we derive a `DALI core' alignment from the original
DALI alignment as follows. We superimpose a given pair structure
according to the DALI alignment, and calculate the distance of each
correspondent residue duad. We remove residue duads with distance
greater than $d_0$, and the aligned segments whose lengths are smaller
than $\rho$. The remaining reduced correspondence is the DALI core of
the original alignment. The transformation optimal to the core is then
determined, and $N_e$ and RMSD of the core are calculated; they are
also listed in the table. It is seen that the performance of CLePAPS is
comparable with DALI. A more detailed comparison on ten protein pairs
is given in the next subsection.

\subsection{Ten `difficult' protein structure pairs}

Ten protein pairs from the Fischer benchmark set of 68 pairs were
regarded as `difficult' for fold recognition, and treated as
a test set by CE and ProSup. The comparison of CLePAPS with
DALI, CE and ProSup is shown in Table 6. Since different
criteria are used there is
no simple direct comparison. For example, a high
RMSD would lead to a large number of equivalent residues.
With an extra restriction in the minimal size for aligned
segments, ProSup usually has a smaller $N_e$ than others.
To make a direct comparison, as done in the last subsection
for DALI alignments, we derive also CE-core alignments from
the original CE alignments. Generally, alignments of CLePAPS
are comparable with those of other alignment tools.

The two proteins of pair 1fxiA: 1ubq are of lengths 96 and 76,
respectively. There is only one member in List-20. The alignment
listed in Table 5 is from the only AFP. This is rather risky,
being prone to local trapping. It is easy for CLePAPS to give
a warning. A simple way of rescue is to use a weaker criterion
for AFPs, e.g. List-12 with $T_{12}=180$ or a lower $T_{20}$.
The alignment listed in Table 6 is obtained by using either
$T_{20}=200$ or $T_{12}=180$.
The CLePAPS alignment for the pair 1bgeB: 2gmfA has nothing in
common with the first of DALI's three alignments. The List-20
of the protein pair has 31 members, but none coincides with
any segment of the first DALI alignment. This means that the local
similarity of the alignment is rather weak. The CLePAPS alignment
for the pair is very similar to the second DALI alignment of $N_e=94$ with
RMSD 3.3\AA. Of the total 82 aligned residue duads, 72 are identical
with those of DALI.

\begin{center}
\parbox{15cm}{Table 6. Comparison of structure alignments
obtained by DALI, CE, ProSup and CLePAPS for `10 difficult'
cases from the Fischer benchmark. $N_e$: total number of
equivalent residue duads; rmsd: RMSD in units of \AA; IDA: number
of residue duads which are identical to those of DALI.\\
}%\vspace{-.8cm}

\begin{tabular}{ll r r r r rr rr}\hline
&&\multicolumn{1}{c}{CE}&\multicolumn{1}{c}{DALI}&
 \multicolumn{1}{c}{CE-core}&\multicolumn{1}{c}{DALI-core}&
 \multicolumn{2}{c}{ProSup} &\multicolumn{2}{c}{CLePAPS}\\
\multicolumn{2}{c}{Pair}&$N_e$/rmsd &$N_e$/rmsd& $N_e$/rmsd&
$N_e$/rmsd& $N_e$/rmsd& IDA& $N_e$/rmsd& IDA\\
\hline
 1fxiA& 1ubq & 64/2.8&  60/2.6&  59/2.5&  55/2.3&  54/2.6&41 &  55/2.4&  42\\ %
 1ten & 3hhrB& 87/1.9&  86/1.9&  85/1.7&  84/1.7&  85/1.7&79 &  84/1.7&  77\\
 3hlaB& 2rhe & 85/3.5&  75/3.0&  71/3.0&  63/2.3&  71/2.7&37 &  65/2.3&  57\\
 2azaA& 1paz & 85/2.9&  81/2.5&  73/2.5&  76/2.1&  82/2.6& 8 &  78/2.3&  72\\
 1cewI& 1molA& 81/2.3&  81/2.3&  78/2.0&  78/1.9&  76/1.9&68 &  78/2.0&  75\\
 1cid & 2rhe & 98/3.0&  97/3.2&  79/2.0&  82/2.0&  84/2.3&70 &  87/2.2&  72\\
 1crl & 1ede &220/3.9& 211/3.5& 155/2.5& 168/2.5& 161/2.6&147& 169/2.7& 146\\
 2sim & 1nsbA&276/3.0& 292/3.3& 236/2.5& 240/2.5& 248/2.6&231& 248/2.6& 213\\
 1bgeB& 2gmfA&109/4.6&  94/3.3&  62/2.7&  79/2.2&  87/2.4&0  &  82/2.4&   0\\ %
 1tie & 4fgf &117/3.0& 114/3.1&  99/2.3&  97/2.2& 101/2.4&48 & 100/2.3&  94\\
\hline
\end{tabular}\\ \end{center}

\subsection{Database search with CLePAPS}

Only 56 of the 68 protein structures in the Fischer benchmark are
distinct. We take 4 proteins from the 68 probes of the benchmark as query
structures: 2mtaC from Class $\alpha$ Fold Cytochrome, 1fxiA from
$\alpha +\beta$ Ubiquitin-like, 1tie from $\beta$ Trefoil, and 3chy
from $\alpha /\beta$ Open sheet. Each
query structure is aligned with each of the 56 target structures
of the benchmark. The first eight structures found by CLePAPS to be similar
to these probes are shown in Table 7. In the benchmark, the
structure counterparts of 2mtaC, 1fxiA, 1tie and 3chy are 1ycc, 1ubq,
4fgf and 2fox, respectively. It is seen that, indeed, 1ycc, 1ubq and
4fgf have the highest $N_e$ among the similar structures found. However,
2fox appears rather behind. A close inspection reveals that the six
structures with a higher $N_e$ than 2fox are rather large, about three
or five times longer than 3chy (except 3adk), and all have a
domain or repeat of the same CATH \cite{cath} topology `Rossmann fold' as 3chy.
Taking the lengths of the aligned proteins into account, the DALI
$Z$-score \cite{holm1998b} assigns the second highest value to 2fox.
The DALI $Z$-scores for alignments to 3chy
are also listed in the table. All the listed structures found for 3chy
are coincident with 3chy, at least at the CATH topology level.

%3chy: alpha/beta Flavodoxin-like CheY-like; cath 3-Layer(aba)-Sandwich
%Rossmann fold. 2fox .. Flavodoxin-related; cath .. Rossmann fold.
%1ede, 2minB*3+up-down, 2liv*2, 1cta, 3adk: T: Rossmann fold.

According to the Fischer benchmark, four probe sequences 1tlk, 3cd4,
3hlaB and 1cid are associated with the structure 2rhe. We have also
taken 2rhe as a query to search the 68 probes for similar folds, and
the results are also listed in Table 7.
As we may expect, the structures of the above four probes do indeed 
appear to be highly ranked among those similar to 2rhe. However,
an additional one, 2fbjL, has the highest value of $N_e$. It is
verified that the two domains of 2fbjL share the same SCOP
superfamily or CATH homology with 2rhe, and 2fbjL has the highest DALI
$Z$-score. Protein 1sacA is not classified as Immunoglobulin (IG) or
IG-like, but it still shares the same CATH Architecture with 2rhe.

%6 IG-like two (1pfc z=4.2, 1ten z=4.8) missing? 2 IG 1fc1A in, 2fbjL (z=17.4)?
%3cd4 12.1, 1cid 8.3, 1tlk 11.3, 1fc1A 7.0, 3hlaB 5.5
%2rhe(114), 1tlk (103), 3cd4 (178), 3hlaB (99) and 1cid (177)
%1ifc 3grs 7rsa 4cpv 3hlaB 6ldh 1shaA 2ccyA 1ycc 2mnr 1hbg 8fabB
%3adk 2hipA 2scpA 2fb4H 2cpp 1ego 4cla 5ptp 1fbpA 1paz 1iqz 2hipA 14
%3cox 1cauA 1lfb 2rhe 2por 2cyp 2minB 2tbvA 4fgf 1rnh 1rbp 1colA
%1gky 1atr 5ptp 20 1fnb 6xia 9rnt 2rhe 28 256bA 4enl 1bovA 2liv 2plv
%1ycc 9 1tca 2gmfA 2ayh 2trxA 1molA 1aozA 1ubq 2gmfA 51 2rhe 28 2fox
%1paz 22 1molA 54 2rhe 28 1ede 1nsbA 3hhrB 4fgf 33 5ptp 20 2trxA 53

\begin{center}
\parbox{15cm}{Table 7. Structures found by CLePAPS as highly similar
to five queries. $Z_{\rm dali}$: the $Z$-score given by
DALI; rmsd: RMSD in \AA units. \\
}%\vspace{-.8cm}

{\setlength{\tabcolsep}{1.2mm}%\small
\begin{tabular}{lr|lr|lr|lrc||lr}\hline
\multicolumn{11}{c}{Query structure}\\
\multicolumn{2}{c|}{2mtaC} &\multicolumn{2}{c|}{1fxiA}
&\multicolumn{2}{c|}{1tie}&\multicolumn{3}{c||}{3chy}
&\multicolumn{2}{c}{2rhe}\\
\hline %\cline{2-11}
ID& $N_e$/rmsd& ID& $N_e$/rmsd& ID& $N_e$/rmsd&
ID& $N_e$/rmsd&$Z_{\rm dali}$& ID& $N_e$/rmsd \\
\hline
1ycc&  71/1.9 &1ubq & 50/2.3 &4fgf &100/2.3 &2liv &102/2.6 &7.9 & 2fbjL &102/1.5\\ %&17.4 1pfc?
1tca&  47/3.1 &1rbp & 48/3.0 &5ptp & 60/2.9 &2minB& 97/2.9 &6.7 & 3cd4  & 89/1.6\\ %&12.1
1hbg&  47/2.6 &5ptp & 44/3.2 &9rnt & 51/2.5 &1tca & 96/2.6 &5.6 & 1cid  & 87/2.2\\ %& 8.3
1atr&  46/3.0 &1aozA& 39/3.1 &1cauA& 49/2.7 &1ede & 88/2.7 &4.9 & 1tlk  & 86/1.7\\ %&11.3
2scpA& 44/2.9 &1nsbA& 38/3.4 &2plv1& 49/2.8 &3adk & 88/3.0 &4.3 & 1fc1A & 76/2.6\\ %& 7.0
2cpp&  43/2.9 &1atr & 35/2.7 &1rbp & 45/2.5 &6ldh & 87/2.7 &6.0 & 1ten  & 68/2.8\\ %& 4.8
4cpv&  42/3.2 &1ifc & 35/2.5 &1fnb & 45/2.9 &2fox & 86/2.8 &6.8 & 1sacA & 67/2.6\\ %& 0.1 1sacA?sandwith
2mnr&  41/2.8 &1fbpA& 35/2.8 &1nsbA& 45/3.4 &1fnb & 84/2.9 &3.1 & 3hlaB & 65/2.3\\ %& 5.5
\hline
\end{tabular}}\\ \end{center}

\subsection{Multiple solutions of alignments}

To be less greedy, CLePAPS generates several alignments from star centers
of highly scored AFPs. Often there is one alignment which has a much
higher $N_e$ than others. (A better measure of significance for
alignments is the $P$-value or $Z$-score.) There are situations where
several meaningful alignments do exist. We see such examples in
the Fischer benchmark.

Proteins 8i1b and 4fgf belong to SCOP fold beta-Trefoil, which exhibits a
three fold rotational symmetry. Indeed, the alignment ranks 1 to 3
clearly demonstrate this symmetry. A more complicated example is the pair
2sim: 1nsbA. Both belong to SCOP fold 6-bladed beta-propeller, but in PFAM
\cite{pfam} classification 2sim has four repeats of PF02012 while 1nsbA
is PF00064. Multiple alignments reflect the rough symmetry. Another
example is the pair 3rubL: 6xia, which contains a SCOP fold TIM
beta/alpha-barrel. Among the multiple alignments, only one follows
a sequential order, while all the others correspond to a circular
permutation. There are, however, non-topological alignments not related 
to a rotational symmetry, which will be described in the next subsection.

Proteins 1osa and 4cpv belong to all alpha SCOP fold EF hand-like. The
former contains two EF hand domains, while the latter has a single domain. The
top two alignments are mainly the two ways of alignments of the EF hand
domains. Another similar example is the pair 3hhrB: 1ten of
SCOP fold IG-like beta-sandwich. 

Proteins 2gbp and 2liv both belong to SCOP class $\alpha /\beta$
superfamily Periplasmic binding protein-like I, consisting of two
CATH 3-Layer ($\alpha\beta\alpha$) Sandwich domains, which will
be denoted by I and II with I near the N-terminus. The two
domains present a repeat of highly
similar $\alpha\beta\alpha$ segments. For example, the AFP search at
length 20 of 2gbp against itself discovers a region of contiguous AFPs
with the correspondence of sites 23 -- 60 to 163 -- 200 around an AFP
of $\sigma = 772$.
The ranks 1 to 3 in CLePAPS alignments of 2gbp to 2liv are mainly the
alignments of domain I to I, II to I, and II to II, respectively.
The AFP search of 2gbp against 2liv ascertains a long
piece of contiguous AFPs with a correspondence of sites 24 -- 80 of 2gbp
to 34 -- 90 of 2liv around the AFP-20 of the highest $\sigma = 804$,
a piece with a correspondence of 167 -- 201 to 37 -- 71 around
an AFP of $\sigma = 733$, and another with a correspondence of
137 -- 181 to 133 -- 177 around an AFP of $\sigma = 606$. These roughly
correspond to the cores of the three alignments, which are of $N_e = 141$ with
RMSD = 2.54\AA, $N_e = 137$ with RMSD = 2.66\AA, and $N_e = 112$ with
RMSD = 2.40\AA, respectively.
A relative move between the two domains with respect to the two
proteins makes the RMSD for the alignment of the proteins
as a whole rather high. This is seen in the DALI
and CE alignments, which are of $N_e= 260$ with RMSD 6.8\AA\ and $N_e =
252$ with RMSD 4.6\AA, respectively.
%The alignments of ranks 1 and 3 are shown in Fig.~?.
%In fact, much finer alignments from subsets of these
%alignments exist, which are those of $N_e = ?$, RMSD = ?\AA and
%$N_e = ?$, RMSD = ?\AA, respectively. This is a characteristic of the
%domain move. (DALI, CE, global alignment at a poor RMSD with $N_e=?$.)

A global alignment involves the spatial arrangement of fragments. A domain
move can destroy a global alignment as a rigid superposition. However, the
alignment of the corresponding sequences of conformational codes is not
affected by the domain move, so it is convenient for discovering 
conservative substructures of domains. A domain move appears in the structure
evolution. It also occurs as conformational change of some flexible
proteins in function. For example, an adenylate kinase (AKE) has a stable
inactive conformation, in addition to an active form, {\it i.e.,} the
open and closed forms. They are represented by structures of 
PDBIDs 4akeA and 1akeA,
respectively. Of course, they have the identical amino acid sequence. Their
code sequences are also highly similar; only three code pairs are of a
negative score. Their (positions, codes and scores) are (K47, $HM$, $-4$),
(I116, $HM$, $-4$) and (V121, $NI$, $-11$), respectively. If we cut the two
structures arround K47, V121 and D159 into four pieces we can align each
piece pair extremely well. The code comparison can easily discover two
cutting points. In order for structure alignment tools based on a rigid
superposition to recognize the four aligned pieces a stringent criterion
for deviation should be imposed for finding multiple solutions.

\subsection{Non-topological alignment and domain shuffling}

It is well known that the 3D structures of two proteins may be
surprisingly similar in secondary structure element packing while
the sequential order of their SSEs is completely different. We do not
see any clear example of such non-topological alignment other
than a simple rotation in the Fischer benchmark. A good example of such
protein pairs is SCOP fold SH3-like barrel 1ihwA: 1sso (of lengths 52
and 62, respectively).\cite{shuffle} We represent an aligned fragment by the
triplet $(a: a', w)$, where $w$ is the fragment length, and $a$
and $a'$ are starting positions in the two proteins. The alignment
of 1ihwA: 1sso consists of 5 pieces:
$$(2: 26; 10), (13: 36; 10), (28: 0; 7), (36: 8; 11), (47: 20; 5).$$
and has $N_e=43$ with RMSD 2.21\AA. The first two segments show that
the N-terminus of 1ihwA aligns with the C-terminus of 1sso while
the other three segments are the alignment between the C-terminus
of 1ihwA and the N-terminus of 1sso. Such alignments cannot be
found by algorithms using dynamic programming.

\section{Discussion}

CLePAPS distinguishes itself from other existing algorithms
for pairwise structure alignment in its use of conformational
letters. The description of 3D segmental structural states
by a few conformational letters aptly balances
precision with simplicity. The substitution matrix CLESUM provides
us with a proper measure of the similarity between these discrete
states or letters. Such a description fits the $\epsilon$-congruent 
problem very well. Furthermore, CLESUM extracted
from the database FSSP of structure alignments contains
information of structure database statistics. For example,
although two frequent helical states are geometrically very
similar, scores between them are relatively low,
which reduces the chance of accidental matching of two irrelevant
helices. The conversion of coordinates of a 3D structure to its
conformational codes requires little computation. Once we
transform 3D structures to 1D sequences of letters, tools
for analyzing ordinary sequences can be applied with some
modification. The use of conformational letters for fast local
similarity search can be integrated into many existing tools to
improve the latter's efficiency.

Recently, a few research groups have developed 
substitution matrices using various coding schemes for
structural alphabets. For example, the coding of Godzik's group
is based on a rigid fragment library, and their matrices were
constructed using multiple structure alignments (HOMSTRAD)
and alignments derived from multiple sequence alignments
(BLOCKS).\cite{godzik} Tyagi {\it et al.} published a matrix
for their alphabet of 16 `protein blocks' representing eight
dihedral angles of pentapeptides.\cite{pb} Also, Chang {\it et
al.} reported a coding based on writhing number defined for
a chain fragment and a matrix generated by using the CLASTALW
alignments of sequences derived from SCOP as
lineages.\cite{writhe} Tung {\it et al.} derived a structural
alphabet by clustering in a plot of two angles of pentapeptides,
and its substitution matrix SASM.\cite{blast3d} Lo {\it et al.}
derived another substitution matrix for their Ramachandran
codes.\cite{lo} Structural
codes convert structures into one-dimensional sequences.
These authors then compared structures by methods of
sequence alignment or database search, but no structure
alignment was considered.

The CLESUM similarity score can be used to sort the importance
of AFPs for a greedy algorithm, such as CLePAPS. Guided by CLESUM 
scores, only the top few structurally similar fragments need to be
examined to determine the superposition for alignment, and hence
a reliable greedy strategy becomes possible. %%%
Since many computational steps are conducted on conformational
codes instead of 3D coordinates it runs faster than other tools,
especially for large proteins. The running time for the 68 pairs
of the Fischer benchmark is less than 2 percent of that of the
downloaded CE local version.

To a certain extent CLePAPS resembles some other
algorithms such as STRUCTAL \cite{structal} and ProSup, but there are 
fundamental differences. Apart from its key feature in its use of 
the representation of conformational letters, CLePAPS is
different from most algorithms in that it never conducts any 
dynamic programming, so it is able to obtain non-topological as
well as topological alignments. However, the conformational
alphabet can be used to find an initial correspondence for 
dynamic programming. We
have tested two ways: the usual Needleman-Wunsch global alignment
method \cite{nw} with some simple penalty for the gaps or
a little more sophisticated way by joining nonoverlapping AFPs,
which is the Needleman-Wunsch alignment of AFPs. A gap penalty is
necessary to keep the aligned pieces less scattered (sequentially).

It should be remarked that suitable tuning of the parameters used 
by CLePAPS is somewhat crucial to its
optimal performance. A large value of basis width $l$ or similarity
threshold $T$ would reduce search times, but at the price of
sensitivity. Our strategy is to use stringent parameters first
for finding reliable seed matches by star tree construction
to initiate the alignment, and then
to fill missing blanks for eventually compensating the sensitivity
loss with relaxed parameters. The `zoom-in' strategy for blank-filling
starts with a vague alignment to avoid local trapping, and then refines
it in later steps. For two remote structures, to keep the size of
List-20 reasonably large a lower $T_{20}$ is preferable. This also
happens when one or both proteins are small. For large proteins a
large $J$ is often necessary for obtaining a star large enough to
avoid local trapping. However, CLePAPS gets warned by a too small degree 
of a star center, so the danger of local trapping can be significantly
diminished. We have tested various combination of parameters. %%%
In general, CLePAPS is not extremely sensitive to the choice of 
parameters. For example, $T_{20}= 300$ or an even lower value may 
be used in most cases to guarantee that the size of List-20 is large 
enough, and $T_8=50$ is often sensitive enough to include
all relevant aligned segments.

The problem of evaluating the significance of a structural alignment by a
$P$-value or $Z$-score is by no means simple. It has been demonstrated
by Lackner et al.\ that optimizing the number of equivalent residues
under a distance cutoff for residue equivalence provides a simple and intuitive
measure of structure similarity.\cite{prosup} The similarity
index SI $={\rm RMSD} \times L_{\min}/N_e$ of Kleywegt and Jones %(1994)
is simple and convenient, where $L_{\min}$ is the shorter length of
the two proteins.\cite{si} The $Z$-score of DALI (with possible modification
in $M(L)$) can be used as another practicable measure of significance.
A detailed discussion of the $P$-value and $Z$-score is beyond the scope 
of this paper.

CLESUM counts only information of
conformation. However, the FSSP alignments from which CLESUM was
derived contain also amino acid information. The use of a
modified CLESUM including also such information would elucidate
biochemical roles in alignment.\cite{zheng-liu}
The idea of using structural alphabets for pairwise structure
alignment is also valid for multiple structure alignment, where
an AFP is replaced by a similar fragment block. An algorithm
has been developed along this line.\cite{lzz}

\begin{quotation}
{This work is supported by the National Natural Science Foundation of
China and National Basic Research Program of China (2007CB814800).}
\end{quotation}

\end{document}